\documentclass[twoside,slac_one]{revtex4}
\usepackage{graphicx}
\usepackage{fancyhdr}
\usepackage{amsmath} % American Mathematics Society standards
\usepackage{bm}% bold math
\usepackage{amsxtra}
\usepackage{amssymb}
\usepackage{amsthm}
\usepackage{latexsym}
\usepackage{lscape}

\begin{document}

\title{Status of the Upsilon Polarization Measurement at CDF}


\author{Niharika Ranjan\footnote{Speaker} (for the CDF collaboration)}
\affiliation{Department of Physics, Purdue University, West Lafayette, IN, USA}

\begin{abstract}
The angular distributions of $\Upsilon(1S) \rightarrow \mu^+ \mu^-$ decays are analyzed using a sample of $\Upsilon(1S)$ mesons in $2.9\;{\rm fb^{-1}}$ of data collected at $\sqrt{s} = 1.96\;{\rm TeV}$.
The results of the one-dimensional angular analysis suggest that $\Upsilon(1S)$ may be longitudinal polarized at high transverse momentum.
This observation is largely inconsistent with NRQCD prediction that predicts transverse polarization at high $p_T$.
%We describe a technique that involves analyzing the complete three-dimensional angular distribution of $\Upsilon(nS) \rightarrow \mu^+ \mu^-$ decays and will be used in future polarization measurements at CDF.
\end{abstract}

\maketitle

\thispagestyle{fancy}


%%%%%%%%%%%%%%%%%%%%%%%%%%%%%%%%%%
\section{Introduction}
The discrepancies between experimental results and theoretical predictions in the field of heavy quarkonium production continue to intrigue our community.
The production cross section and polarization of heavy quarkonia are important observables that are predicted by the models used to describe heavy vector meson production. 
It is essential to measure these observables in order to test the theoretical models.

The Color Singlet Model (CSM) based on QCD was the first model that was used to describe the production mechanism of heavy quarkonia~\cite{bib-csm1,bib-csm2}.
%In the CSM, heavy vector meson production is predicted to be suppressed.
The CDF Run-I measurements of prompt $J/\psi$ and $\psi$(2S) production cross section showed a significant excess compared to the predictions of the CSM~\cite{bib-jpsi}.
This lead to the development of the Color Octet Model (COM) based on NRQCD, which includes Feynman diagrams for color octet $Q\bar{Q}$ states in addition to the color singlet $Q\bar{Q}$ states, prior to formation of a quarkonium state.
Although the measurement of $J/\psi$ production cross section~\cite{bib-com1} seemed to agree well with the predictions of the octet model, it is important to note that the parameters in COM were tuned to match the normalization in data.
Hence, in order to validate the COM it was important to measure the polarization of heavy vector mesons, which is independent of the cross section measurement.
The COM predicts that heavy vector mesons should be transverse polarized at high momentum~\cite{bib-com2}.
However, CDF Run-I~\cite{bib-jpsi-run1} and Run-II~\cite{bib-jpsi-run2} measurements of $J/\psi$ polarization showed a trend towards longitudinal polarization at high momentum in contrast to the predictions of the COM.
This observation lead to the question that perhaps the charm quark is too light in order to make reliable predictions using NRQCD.
Thus it is essential to measure polarization of the bottomonium states that are bound states of the heavier bottom quark and compare the result with NRQCD predictions.

In these proceedings, we present the results of the measurement of $\Upsilon$(1S) polarization using $\Upsilon \rightarrow \mu^+ \mu^-$ decays in $2.9\;{\rm fb^{-1}}$ of CDF data.
The analysis technique is essentially similar to the technique used in the CDF Run-I measurement~\cite{bib-upsilon-run1}.

%%%%%%%%%%%%%%%%%%%%%%%%%%%%%%%%%%
\section{Measurement of Spin Alignment or Polarization}
Although the nomenclature used for the measured parameter is {\it polarization}, we essentially measure the spin alignment of the vector mesons with respect to a reference axis.
Conventionally, the S-channel helicity frame is used for the polarization measurement at hadron colliders, in which the reference axis is defined by the $\Upsilon$ momentum in the lab frame.
The angular distribution of the $\Upsilon$ decay products is given by:
\begin{equation}
\frac{d\Gamma}{dcos\theta^*} \propto 1 + \alpha ~cos^2\theta^*
\label{eq-angdist}
\end{equation}
where $\theta^*$ is the angle between the momentum of $\mu^+$ in the $\Upsilon$ rest frame and the $\Upsilon$ boost direction.
The parameter $\alpha$ is defined as, $\alpha = (\sigma_T - 2\sigma_L)/(\sigma_T + 2\sigma_L)$ where $\sigma_T$ and $\sigma_L$ are the production cross sections for purely transverse and purely longitudinal polarized vector mesons respectively.
The $\Upsilon$ mesons are said to be transverse polarized if $\alpha = +1$, longitudinal polarized if $\alpha = -1$ and unpolarized if $\alpha$ is measured to be zero.

%%%%%%%%%%%%%%%%%%%%%%%%%%%%%%%%%%
\section{Data Sample}
The data sample used for the measurement corresponds to an integrated luminosity of $2.9\;{\rm fb^{-1}}$ of $p\bar{p}$ collisions at a center of mass energy of $\sqrt{s} = 1.96\;{\rm TeV}$. 
The $\Upsilon$ candidates are selected using the dimuon trigger that requires a pair of oppositely charged muons with transverse momentum of $p_T(\mu) > 3\;{\rm GeV}$.
The trigger algorithm basically requires loose matching between charged particle tracks in the Central Outer Tracker (drift chamber) and muon hit segments in the central muon detectors that have pseudorapidity coverage of $|\eta| < 0.6$~\cite{bib-cdf}.
In addition to the trigger $p_T(\mu)$ criteria, one of the muon is required to have transverse momentum of at least $4\;{\rm GeV}$ in the offline selection process.
Additional track quality, trigger acceptance, dimuon vertex probability and decay length ($L_{xy}$) cuts are imposed offline in the analysis.
Furthermore, the $\Upsilon$ candidates are required to have transverse momentum in the range $2 < p_T < 40\;{\rm GeV}$ and rapidity in the region $|y| < 0.6$.

\begin{figure}[ht]
\centering
\includegraphics[width=80mm]{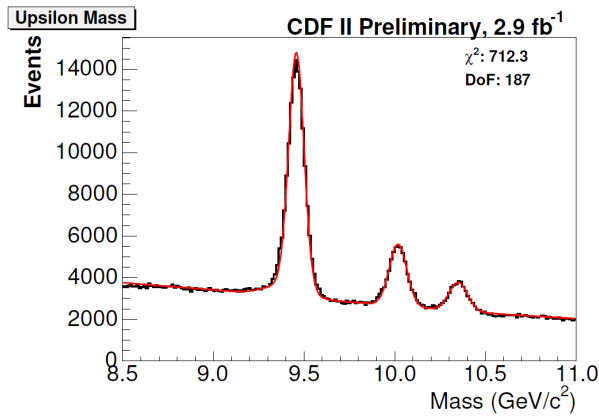}
\caption{Invariant mass distribution of the dimuon candidates using $2.9\;{\rm fb^{-1}}$ of data. The $\Upsilon$(1S) yield is $\approx 83,000$.} 
\label{fig-dimuon}
\end{figure}

The invariant mass distribution of the selected dimuon candidates is shown in Figure~\ref{fig-dimuon}.
Although the figure includes the $\Upsilon$(1S) along with the higher $\Upsilon$(nS) states, we focus here on measuring polarization of the $\Upsilon$(1S) state.
%The signal shape is obtained by studying Monte Carlo samples and the shape of the background is derived from the sideband regions.
The signal region is defined within $2.5\sigma$ from the mean of the $\Upsilon$(1S) mass peak.

%%%%%%%%%%%%%%%%%%%%%%%%%%%%%%%%%%
\section{Measurement of Polarization Parameter}
The polarization parameter $\alpha$ can be determined experimentally by measuring the distribution of cos$\theta^*$ in data.
The sample is divided in eight bins of $\Upsilon$ transverse momentum and the distribution of $d\Gamma/dcos\theta^*$ is measured within each $p_T(\Upsilon)$ bin.
Since the angular distribution has a quadratic dependence on cos$\theta^*$ as shown in Equation~\ref{eq-angdist}, the distribution is symmetric about zero and consequently, the measurement can be performed as a function of $|$cos$\theta^*|$.
Within each $p_T(\Upsilon)$ bin, the angular distribution is obtained by measuring the $\Upsilon$ yield in tens bins of $|$cos$\theta^*|$.

\begin{figure}[ht]
\centering
\includegraphics[width=120mm]{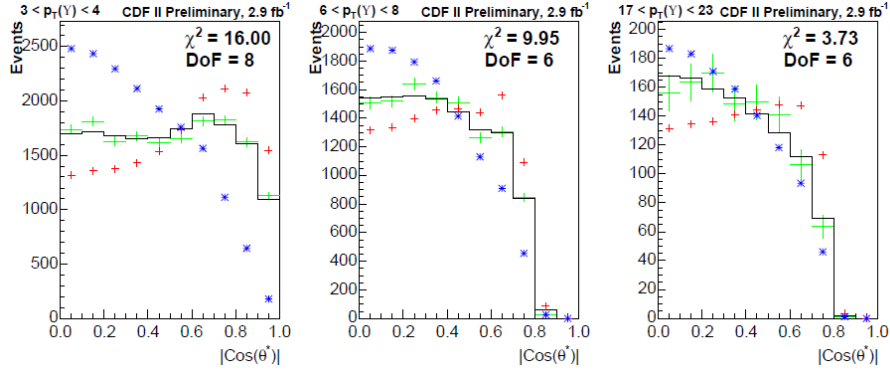}
\caption{Angular distribution and $\chi^2$ fit projections in three momentum bins. The green crosses show the data and black histograms show the best fit results. The blue and red points depict the scaled templates for purely longitudinal and purely transverse polarized vector mesons obtained from Monte Carlo samples.}
\label{fig-etafit}
\end{figure}

The signal component in the $|$cos$\theta^*|$ distribution is described by the weighted sum of two templates, one for purely transverse ($T_T$) and one corresponding to purely longitudinal ($T_L$) polarized vector mesons.
The $T_T/T_L$ templates obtained from Monte Carlo simulations are corrected for detector acceptance and trigger efficiency effects. 
The angular distribution of the background component is obtained from the mass sideband regions.
A $\chi^2$ fit is performed in which the signal component is described using the function:
\begin{equation}
\eta ~T_L + (1-\eta) ~T_T
\label{eq-chi2func}
\end{equation}
where $\eta$ is related to the polarization parameter $\alpha$:
\begin{equation}
\eta  = \frac{1-\alpha}{3 + \alpha}
\label{eq-eta-alpha}
\end{equation}
The shapes of the two Monte Carlo templates $T_T/T_L$ are fixed in the $\chi^2$ fit.
The parameters describing the shape of the background component are set free in the fit but are constrained by simultaneously fitting the invariant mass sideband regions.
Using this fitting procedure, we obtain the value of $\alpha$ by measuring $\eta$ within each $p_T(\Upsilon)$ bin.
Figure~\ref{fig-etafit} shows the result of the $\chi^2$ fit for three of the eight $p_T(\Upsilon)$ bins defined in the analysis.

We investigate several potential sources of systematic effects such as the trigger efficiency parameterization used in Monte Carlo, the $p_T$ re-weighting technique use on Monte Carlo samples, the $\Upsilon$ invariant mass parameterization, the width of the signal range, minimum $p_T(\mu)$ requirement, $|$cos$\theta^*|$ bin width, and number of $|$cos$\theta^*|$ bins.
The uncertainty on the trigger efficiency function measured on data introduces a very small systematic uncertainty of $0.007$ on the polarization parameter.
Systematic uncertainties due to the remaining sources are found to be negligible.
The total uncertainty on the measurement of $\alpha$ is thought to be dominated by statistical errors.

%%%%%%%%%%%%%%%%%%%%%%%%%%%%%%%%%%
\section{Results}
\begin{figure}[ht]
\centering
\includegraphics[width=120mm]{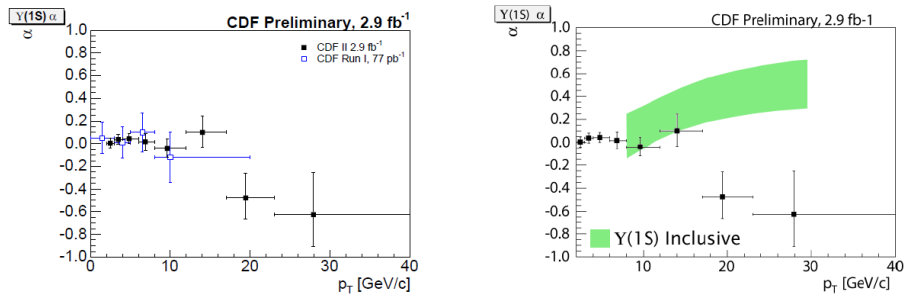}
\caption{Distribution of polarization parameter $\alpha$ as a function of $p_T(\Upsilon)$ for the $\Upsilon$(1S) state. The plot on the left shows a comparison with the CDF Run-I result. The green band in the plot on the right shows the prediction of NRQCD. The width of the band results from the uncertainty in feed down from higher excited states.}
\label{fig-alpha}
\end{figure}

The distribution of measured polarization parameter $\alpha$ as a function of $p_T(\Upsilon)$ for the $\Upsilon$(1S) state is shown in Figure~\ref{fig-alpha}. 
We measure almost zero polarization at low momentum, which is consistent with the previous Run-I measurement.
The measurement shows a trend towards longitudinal polarization at high momentum that disfavors the NRQCD prediction~\cite{bib-com2} of transverse polarization at high $p_T$.

%%%%%%%%%%%%%%%%%%%%%%%%%%%%%%%%%%
\section{Analyzing the Complete Angular Distribution}
In the future, we plan to perform the analysis of the complete three-dimensional angular distribution for a heavy vector meson decaying to fermions that can be written as:
\begin{equation}
\frac{d\Gamma}{d\Omega} = 1 + \lambda_{\theta} cos^2\theta + \lambda_{\varphi} sin^2\theta cos2\varphi + \lambda_{\theta\varphi} sin2\theta cos\varphi
\label{eq-fullangdist}
\end{equation}
In the polarization analyses conducted so far, only the $\lambda_{\theta}$ (equivalent to $\alpha$) parameter has been measured in the S-channel helicity frame.
Recently it has been emphasized~\cite{bib-3d} that in order to fully understand $\Upsilon$ polarization, all the three coefficient in the angular distribution, $\lambda_{\theta}$, $\lambda_{\varphi}$ and $\lambda_{\theta\varphi}$ should be measured simultaneously.
Furthermore, it is suggested that the Collins-Soper frame is a more natural choice for comparison with theoretical predictions.
As a consistency check, we intend to measure a frame-invariant quantity $\tilde{\lambda} = (\lambda_{\theta} + 3 \lambda_{\varphi})/(1 - \lambda_{\varphi})$ derived from the parameterization of the angular distribution measured in each reference frame used in the analysis.

%%%%%%%%%%%%%%%%%%%%%%%%%%%%%%%%%%
\section{Summary}
We presented the latest measurement of $\Upsilon$(1S) polarization using $2.9\;{\rm fb^{-1}}$ of data.
The analysis was conducted using the conventional technique that involves measuring only one of the three parameters in the angular distribution of $\Upsilon$(1S) decays, namely $\lambda_{\theta}$ or $\alpha$.
The results of the measurement suggested that the $\Upsilon$(1S) are longitudinal polarized at high $p_T$, which is inconsistent with NRQCD prediction.
In the future polarization measurements at CDF, we plan to conduct a more comprehensive study of the full three-dimensional angular distribution.
The new technique involves measuring all the three coefficients in the angular distribution of all three $\Upsilon$(nS) states.

%%%%%%%%%%%%%%%%%%%%%%%%%%%%%%%%%%
\begin{acknowledgments}
We are grateful to the Fermilab staff and the technical staffs of the participating institutions for their vital contributions. The work presented in these proceedings was supported by the U.S. Department of Energy and National Science Foundation; the Italian Istituto Nazionale di Fisica Nucleare; the Ministry of Education, Culture, Sports, Science and Technology of Japan; the Natural Sciences and Engineering Research Council of Canada; the National Science Council of the Republic of China; the Swiss National Science Foundation; the A.P. Sloan Foundation; the Bundesministerium f$\ddot{u}$r Bildung und Forschung, Germany; the Korean World Class University Program, the National Research Foundation of Korea; the Science and Technology Facilities Council and the Royal Society, UK; the Institut National de Physique Nucleaire et Physique des Particules/CNRS; the Russian Foundation for Basic Research; the Ministerio de Ciencia e Innovaci\'{o}n, and Programa Consolider-Ingenio 2010, Spain; the Slovak R$\&$D Agency; the Academy of Finland; and the Australian Research Council(ARC).
\end{acknowledgments}

\bigskip % extra skip inserted

\end{document}